\documentstyle[aps,prl]{revtex}
\begin{document}
\title{The Gravitational Constant, the Chandrasekhar Limit, and
  Neutron Star Masses}
\author{S. E. Thorsett} 
\address{Joseph Henry Laboratories and
  Department of Physics, Princeton University, Princeton, NJ 08544}
\maketitle
\begin{abstract}
  The Chandrasekhar mass limit sets the scale for the late
  evolutionary stages of massive stars, including the formation of
  neutron stars in core collapse supernovae.  Because its value
  depends on the gravitational constant $G$, the masses of these
  neutron stars retain a record of past values of $G$. Using Bayesian
  statistical techniques, I show that measurements of the masses of
  young and old neutron stars in pulsar binaries limit $\dot
  G/G=(-0.6\pm2.0)\times10^{-12}\mbox{yr}^{-1}$ (68\% confidence) or
  $\dot G/G=(-0.6\pm4.2)\times10^{-12}\mbox{yr}^{-1}$ (95\%
  confidence).
\end{abstract}
\pacs{04.40.-y, 97.10.Nf, 97.60.Jd}


\paragraph{Introduction}

Whether or not the fundamental ``constants'' of nature vary with time
has been a question of considerable interest since Dirac suggested
that the gravitational force may be weakening with the expansion of
the universe \cite{dir37}.  Although general relativity predicts $\dot
G$ is identically zero, a variable $G$ is expected in theories such as
the Brans-Dicke scalar-tensor theory and its extensions
\cite{wei72,wil93,dgg90,de92}, and has recently received renewed
attention in the context of extended inflationary cosmology
\cite{ls89b}.

The most direct experimental limits on $\dot G/G$ come from monitoring
the separations of orbiting bodies, since from Kepler's laws it is
easily shown that $\dot G/G=-\dot a/a$, where $a$ is the orbital
semimajor axis.  Early studies based on ancient occultation and
eclipse observations had relatively low precision
\cite{fot20,dic66,mul78}, but strong limits are now available from
direct measurements. For convenience, I define $\zeta_{-12}\equiv(\dot
G/G)/10^{-12}\mbox{yr}^{-1}$. Lunar laser ranging experiments yield
$\zeta_{-12}=0\pm11$ \cite{mssr91}, while radar ranging to {\it
  Viking} gives $\zeta_{-12} = 2\pm4$ \cite{haa+83} or $-2\pm10$
\cite{sha90b}, depending on assumptions about solar system mass
uncertainties and correlations between model parameters.  Similarly,
observations of the pulsar--white-dwarf binary PSR\,B1855+09 yield
$\zeta_{-12}=-9\pm18$ \cite{ktr94}.  (Observations of the double
neutron-star binary PSR\,B1913+16 give $\zeta_{-12}=4\pm5$
\cite{tay93a}, but this limit is greatly weakened when the
$\dot{G}$-driven variation in the gravitational self-energy of the
stars is considered \cite{nor90}.)

Indirect evidence about past values of $G$ can be obtained from
comparison of big-bang nucleosynthesis models with the observed $^4$He
abundance \cite{bar78}. A recent reanalysis argues that $0.7G_0<G_{\rm
  BBN}<1.4G_0$ \cite{akr90}, corresponding to
$|\zeta_{-12}|\lesssim0.9$ for a power law variation of $G$, or
$|\zeta_{-12}|\lesssim40$ for a linear variation of $G$.  Limits are
also obtained from considerations of the long term stability of
clusters of galaxies and globular clusters \cite{ds74}
($|\zeta_{-12}|\lesssim40-60$), or from evolution of the Sun or other
stars, since the luminosity of a star $L\propto G^7$
\cite{tel48,dic62}. Unfortunately, the Earth preserves only a crude
memory of the early luminosity of the Sun, so paleontological evidence
gives only a weak limit, $|\zeta_{-12}|<100$ \cite{cs76}.

Fortunately, a much more precise record of early stellar evolution can
be found in the galactic population of neutron stars, whose masses are
set at their time of formation by the balance between the Fermi
degeneracy pressure of a cold electron gas and the gravitational
force, through the Chandrasekhar limit \cite{cha31,wei72}:
\begin{equation}
M_{\rm ch}\sim
\left(\frac{\hbar^{3/2}c^{3/2}}{G^{3/2}m_N^2}\right),
\end{equation}
where $m_N$ is
the mass of the neutron.  Because $M_{\rm ch}$ sets the mass scale in
the late stages of stellar evolution \cite{tww96}, we expect the
average neutron star mass $\mu\sim M_{\rm ch}$, which implies $\dot G/G
=-2\dot\mu/3\mu$. In this Letter, I show how
observations of neutron star masses and ages can be used to set tight
limits on $\dot\mu$ and hence $\dot G$.

\paragraph{The Masses and Ages of Neutron Stars}

There are now five double neutron star binaries known.  In each case,
five Keplerian parameters can be very precisely measured by pulse
timing techniques \cite{mt77}: the binary period $P_b$, the the
projection of the orbital semimajor axis on the line of sight
$x\equiv a_1\sin i$, the eccentricity $e$, and the time and longitude
of periastron, $T_0$ and $\omega_0$. These parameters are related to
the pulsar and companion masses, $m_1$ and $m_2$, through the mass
function
\begin{equation}
\label{massfunc}
f=\frac{(m_2\sin
  i)^3}{(m_1+m_2)^2}=\frac{4\pi^2x^3}{GP_b^2}.
\end{equation}
In each case, the relativistic advance of the angle of periastron,
$\dot\omega$, has also been measured, which yields an estimate of the
total system mass $m_t=m_1+m_2$. For three systems (PSRs~B1534+12,
B1913+16, and B2127+11C) the measurement of the combined effects of
the transverse Doppler shift and the gravitational redshift allow the
individual determination of the pulsar and companion masses. In
Table~1, I collect the measurements of $f$, $m_1$, $m_2$, and $m_t$,
and their uncertainties (the uncertainty in $f$ can be neglected). To
a good approximation, the measurements are independent and normally
distributed, so if $\hat m_i$ is the true value of one of $m_1$,
$m_2$, or $m_t$, then the probability that we will measure $m_i$ is
\begin{equation}
\label{obserr}
P(m_i|\hat
m_i)=\frac{1}{\sqrt{2\pi\sigma_i^2}}e^{-(m_i-\hat m_i)^2/2\sigma_i^2}.
\end{equation}

An upper limit to the age of a pulsar can be determined from the rate
at which the pulsar period $P$ is increasing because of the loss of
rotational energy to radiation, $t_c=P/2\dot{P}$.  The progenitors of
the double neutron star systems were binary systems consisting of two
massive stars that underwent successive supernova explosions.  Because
the lifetime of a neutron star progenitor is only about $10^7$\,yrs,
the age difference between the pulsar and companion can be
neglected. In the case of PSR~J1518+4904, for which $t_c$ is at least
16\,Gyr, I assume instead an upper age limit of 10\,Gyr, about the
age of the Galactic disk.

PSR~B2127+11C is a special case, because it is in the globular cluster
NGC\,7078 (M15), where interactions can result in new companions being
exchanged into binaries. Indeed, in the standard scenario, neutron
star formation is completed within the first $\sim10^7$\,yrs of the
cluster lifetime, so the small characteristic age of the pulsar is
understood as the time since the pulsar was (last) spun-up by a
companion.  In this case, the relevant age for both neutron stars is
the cluster age, which can be found by comparing stellar structure
models with observations of the color-magnitude distribution of
cluster members.  The most recent calculations, using the latest
nuclear equation of state and opacity data, find a cluster age of
12-13\,Gyr \cite{sdw96}, somewhat younger than previous estimates
\cite{scs93,ck95}.

\paragraph{Variability of the Average Neutron Star Mass}

In Figure~1, I display the available information on the ages and
masses of neutron stars in double neutron star binaries. There is
certainly no evidence that the average mass has changed by more than a
few tenths of a solar mass in the last 12~Gyr. I introduce a model
in which the average mass varies as $\mu=\mu_0-\dot\mu t$, where $t$
is the neutron star age, with the goal of estimating the posterier
density $P(\dot\mu|\{x_i\},\{t_i\})$, where $\{x_i\}$ and $\{t_i\}$
are the observations of neutron star masses and ages.

The underlying distribution of neutron star masses is unknown, but the
tight clustering of masses of young stars, for which $\dot G$ is
unimportant, suggest that a normal distribution with variance $s$
about $\mu$ is reasonable.  Hence the probability that a neutron star
of age $\hat t$ will have mass $\hat m$ is
\begin{equation}
P(\hat{m}|\hat t, \mu_0, \dot\mu, s)=
\frac{1}{\sqrt{2\pi s}}e^{-(\hat m-\mu_0-\dot\mu\hat t)^2/2s}.
\label{nseqn}
\end{equation}
I consider a normal distribution because the present data set is too
small to justify a more complex model; other distributions (e.g.,
uniform between an upper and lower bound \cite{fin94}) are possible,
but the final results are fairly insensitive to the form of
Eq.~(\ref{nseqn}).

Assuming a uniform prior density for $\dot\mu$, I use Bayes'
Theorem to write $P(\dot\mu|\{x_i\}, \{t_i\})\propto
P(\{x_i\}|\dot\mu,\{t_i\})$, where the proportionality constant is set
by the normalization condition $\int
P(\dot\mu|\{x_i\},\{t_i\})d\/\dot\mu=1$. Then
\begin{eqnarray}
P(\{x_i\}|&&\dot\mu,\{t_i\})\nonumber\\ && = \int\int
P(\{x_i\}|\dot\mu,\mu_0,s,\{t_i\})\,\pi(\mu_0)\,\pi(s)\,d\mu_0\,ds,
\label{museqn}
\end{eqnarray}
where I take the prior density $\pi(\mu_0)$ to be uniform for positive
$\mu_0$, and $\pi(s)$ as uniform in $\log s$ \cite{edj+71}.

To evaluate Eq.~(\ref{museqn}), note that the independence of the
measurements of the five binary systems allows the factorization
\widetext
\begin{equation}
\label{factoreqn}
P(\{x_i\}|\dot\mu,\mu_0,s,\{t_i\})=\prod_i P(x_i|\dot\mu,\mu_0,s,t_i)
=\prod_i\int P(x_i|\dot\mu,\mu_0,s,\hat t_i)\,P(\hat t_i|t_i)\,d\hat t_i,
\end{equation}
where now $x_i=\{m_2, m_t\}$ for PSRs~B1534+12, B1913+16, and
B2127+11C, and $x_i=\{f, m_t\}$ for PSRs~J1518+4904 and B2303+46.  I
take $P(\hat t_i|t_i)$ uniform for $\hat t_i<t_i$ and zero otherwise,
except for PSR~J1518+4904, where I make the further
(conservative) assumption that $P(\hat t_i|t_i)=0$ for $\hat
t_i>10$\,Gyr, and for PSR~B2127+11C, for which I take $P(\hat
t_i|t_i)=\delta(\hat t_i-12\mbox{Gyr})$. 
We can further factor
\widetext
\begin{equation}
P(m_2, m_t |\dot\mu,\mu_0,s,t_i)=\int\int P(m_2|\hat m_2)\,P(m_t|\hat
m_t)\,P(\hat m_2|\dot\mu,\mu_0,s,\hat t)\,P(\hat m_t-\hat
m_c|\dot\mu,\mu_0,s,\hat t)\,d\hat m_t\,d\hat m_2,
\end{equation}
and a similar, more complex expression for $P(f, m_t
|\dot\mu,\mu,s,t_i)$, using Eq.~(\ref{massfunc}) and $P(\cos i)=1/2$.
Using Eqs.~(\ref{obserr}) and (\ref{nseqn}), we can then (numerically)
evaluate $P(\dot\mu|\{x_i\},\{t_i\})$. I find
$\dot\mu=-1.2\pm4.0(\pm8.5)\times10^{-3}
M_\odot\mbox{Gyr}^{-1}$ at the 68\% (95\%)
confidence level, corresponding to $\dot
G/G=-0.6\pm2.0(\pm4.2)\times10^{-12}$.

\paragraph{Discussion}

The measurement $\zeta_{-12}=-0.6\pm2.0$ is a factor five tighter than
earlier limits. It is important to understand how sensitive this
result is to our model assumptions, the most critical of which concern
the neutron star ages.  Independent evidence that pulsars with small
spin-down rates are old comes from studies of pulsars in binaries with
cooling white dwarf companions. For example, an upper limit on the
optical luminosity of the companion of PSR~B1855+09 yields a minimum
system age of 4\,Gyr, comparable to the pulsar timing age $t_c=5$\,Gyr
\cite{kdk91,ktr94}.

Some post-formation mass transfer is required to recycle the pulsars
to their observed periods, however a very small transfer ($ll
0.1M_\odot$) is sufficient in all cases (and the physics of spin-up is
presumed independent of time).  Of more concern is the age of
PSR~B2127+11C, which is based on the standard model of neutron star
formation: core collapse in massive stars. It has been proposed that
some neutron stars in globular clusters may result from accretion
induced collapse (AIC) of O-Ne-Mg white dwarfs \cite{nom87,bv91},
though theoretical and observational uncertainties remain \cite{bv91}.
In the most unfavorable case, both PSR~B2127+11C and its companion
were (separately) formed by AIC before exchanging into the current
binary.  The pulsar may then be young, with $\hat t_i<t_c$, while the
companion could have any age less than the cluster age. Using these
assumptions, I find the weaker limit $\zeta_{-12}=2.3\pm5.0$.

Mass determinations of pulsars such as PSR~B1855+09, for which white
dwarf cooling ages are available, are clearly of great interest. In
fact, observations of the Shapiro time delay in this system yield
$m_1=1.50^{+0.26}_{-0.14}M_\odot$ \cite{ktr94}, but the uncertainties
are still too large for this measurement to contribute significantly
to our estimate of $\zeta_{-12}$.

Time variability of the Chandrasekhar limit has other potentially
observable implications, most notably for Type Ia supernovae, which
are widely believed to be the thermonuclear disruptions of
Chandrasekhar mass white dwarfs \cite{ww86}, and which are remarkable
standard candles, with an intrinsic dispersion of only 0.12~magnitudes
when multicolor light curve shape corrections are done \cite{rpk96}.
The optical emission is due to decay of $^{56}$Ni.  In a naive model,
$^{56}$Ni production will be roughly proportional to mass, and hence
to $G^{-3/2}$. If $G$ is proportional to the expansion parameter
$R^{\sigma}$, then $\dot G/G=\sigma H_0$. An ambitious program on
existing telescopes could measure the average luminosity of supernovae
at $z=1$ to $\sim 0.05$\,magnitudes \cite{gp95}, corresponding (for
$H_0=60\mbox{km\,s}^{-1}\mbox{Mpc}^{-1}$) to the magnitude change
produced by $\zeta_{-12}=2.7$.

\acknowledgments I thank R. W. Sayer, R. J. Dewey, D. J. Nice and P.
J. E. Peebles for interesting conversations, and J. H. Taylor,
R. V. Wagoner, and an anonymous referee for
comments on the manuscript.


\begin{figure}
\caption{Masses of the neutron stars in the five binaries of
  Table~1. Circles indicate individual stars; squares are the average
  mass in cases where the individual masses cannot be determined. Ages
  shown are upper limits, except for PSR~B2127+11C, as described in
  text. Ages of the components of a binary are offset slightly for
  clarity; uncertainties in the mass of a pulsar and its companion are
  not independent. The variation in the average neutron star mass
  corresponding to $\zeta_{-12}=\pm 10$ is shown.}
\end{figure}

\begin{table}
\begin{tabular}[h]{llllll}
& Mass      & Pulsar    & Mass      & Spindown & Ref \\
&           & mass      & function  & age $t_c$ \\
& $M_\odot$ & $M_\odot$ & $M_\odot$ & (Gyr) \\
\hline 
J1518+4904 & 2.65(7) & --- & 0.1159876 & $>16$ & \protect{\cite{say96}} \\
B1534+12 & 2.67835(16) & 1.338(12) & --- & $0.25$ & \protect{\cite{arz95}} \\
B1913+16 & 2.82843(2) & 1.442(3) & --- & $0.11$ & \protect{\cite{tay92}} \\
B2127+11C & 2.7121(6) & 1.349(40) & --- & $0.10$ & \protect{\cite{dk96}} \\
B2303+46 & 2.60(6) & --- & 0.2462832 & $0.03$ & \protect{\cite{arz95}} \\
\end{tabular}
\end{table}

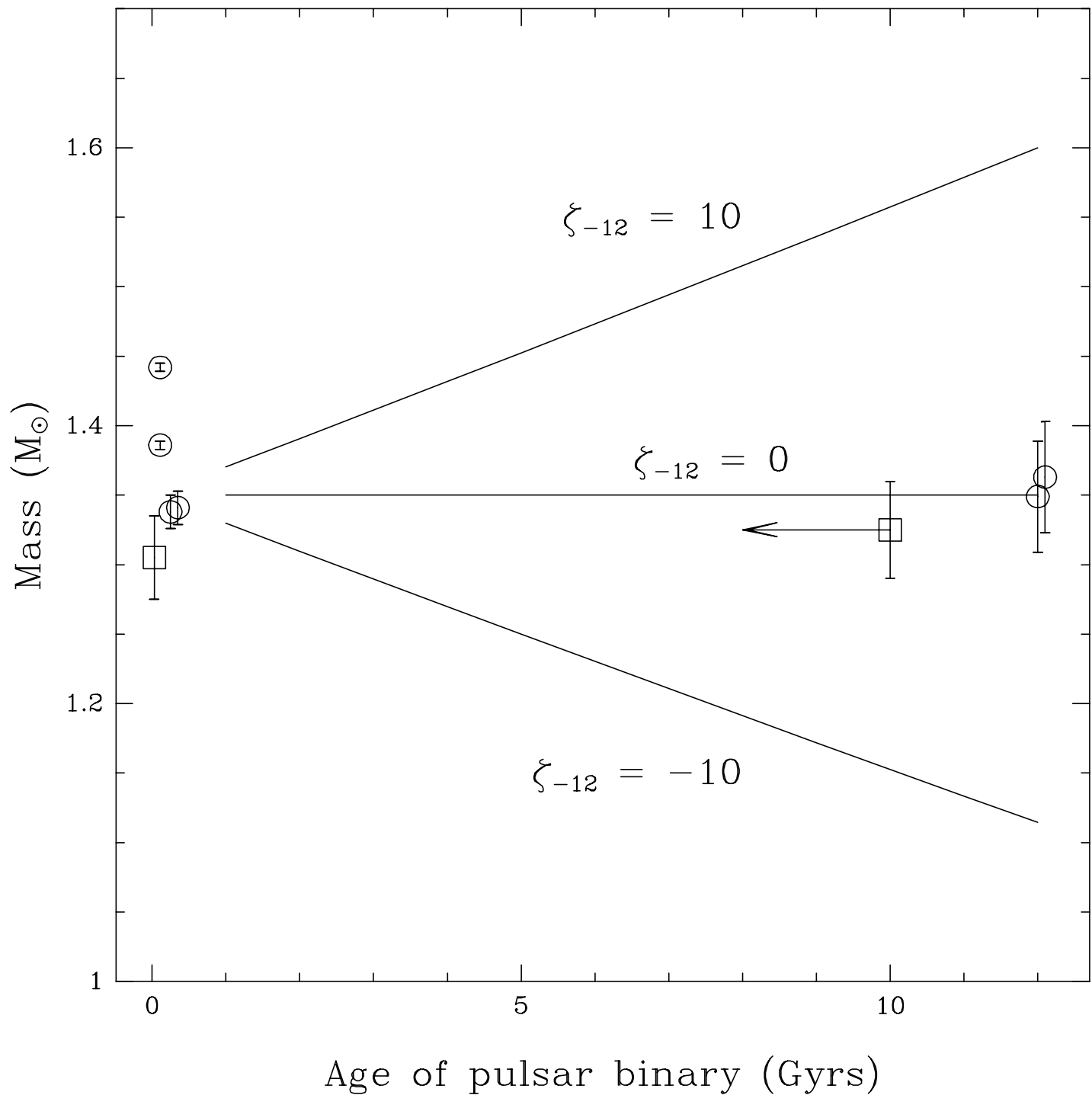
\end{document}